# STEM as Public Computation and Boundary Play

Pratim Sengupta & Marie-Claire Shanahan
University of Calgary - Werklund School of Education
pratim.sengupta@ucalgary.ca
mcshanah@ucalgary.ca

*Abstract:* In this paper, we introduce "public computation" as a genre of learning environments that can be used to radically broaden public participation in authentic, computation-enabled STEM disciplinary practices. Our paradigmatic approach utilizes open source software designed for professional scientists, engineers and digital artists, and situates them in an undiluted form, alongside live and archived expert support, in a public space. We present a case study of DigiPlay, a prototypical public computation space we designed at the University of Calgary, where users can interact directly with scientific simulations as well as the underlying open source code using an array of massive multi-touch screens. We argue that in such a space, public interactions with the code can be thought of as "boundary work and play", through which public participation becomes legitimate scientific act, as the public engages in scientific creation through truly open-ended explorations with the code.

## 1. Introduction

Heidegger famously remarked that the essence of technology is nothing technological (Heidegger, 1993). The eminent Canadian scientist and public philosopher Ursula Franklin argued that technology can be best understood not as a set of tools but as a contextually embedded practice (Franklin, 1990). This perspective implies that technology should be viewed not only as ways and means of performing disciplinary work, but also in light of broader norms of participation in disciplinary and ancillary cultures that develop around localized technological infrastructure. For example, while the broad practices of programming can be explained in terms of generalized computational abstractions and algorithmic dexterity (Wing, 2006), professionals in specific disciplines often require and develop specialized programming tools and localized practices suited for their disciplinary and/or institutional goals (Schmidt, 2006). In Heideggerian terms, this corresponds to the "frame" around technology, which, according to Heidegger, is the essence of technology (Heidegger, 1993).

In this paper, we paradigmatically argue for a frame shift in the technological infrastructure as it pertains to computationally intensive STEM and public education. We introduce *public computing* as a new form of open-ended, public learning environments, in which visitors can directly access, modify and create complex and authentic scientific work through interacting with open source computing platforms. Building on Shanahan, Burke and Francis (2016), we adopt the position that authentic encounters with integrated STEM, for experts and beginners alike, involve the experience of multi-, inter- or transdisciplinarity. The emphasis on "integration" implies that the diverse, individual STEM fields of knowledge and practice should be merged in a manner that reveals big ideas and representational practices that unify or transcend specific disciplines (Nathan et al., 2013; Berland, 2013; Sengupta, Krishnan, Wright & Ghassoul, 2015). In formal education, the meaning of individual STEM disciplines is often formed in reference or opposition to codified forms of disciplinary knowledge and culture such as curricula, textbooks and accepted teaching and learning models. Public STEM spaces, however, can be undocked from those codified meanings offering places, technological means and human capital to explore new ways of knowing and being, offering opportunities to play with disciplinary meanings and expertise in authentic, yet novel and unexpected ways. Open source computing can further facilitate this process by opening up the "code", which often reifies epistemic and representational work of experts, for the public. The opening up of epistemic and representational possibilities, we argue, are both due the structural affordances of the computing media (e.g., open source and touch-based interactivity), as well as the opportunities of collaboration with friends, strangers and experts that often get taken up through joint action, as users configure and reconfigure novel scientific representations and their explanations.

To this end, we present a paradigmatic case study, where we introduce DigiPlay, a public learning environment that uses open source computing for STEM experiences. We present the theoretical underpinnings of our work, and then present a qualitative analysis of participants' experiences that highlight how learning in such a space can be understood as boundary work and play through the construction and re-configuration of figured worlds.

## 2. STEM as Figured Worlds and Boundaries

When conceptualizing science and technology as human practices, the idea of figured worlds is salient. School science, for instance, is in part the constant re/creation by students and teacher of what science is, what its practices are and what types of people are and can be a part of that practice (e.g., Carlone, Scott & Lowder, 2014). The world of school science is what Holland, Skinner, Lachicotte and Cain (2001) would describe as an 'as if' world, where participants become part of acting and creating a shared set rules, norms and values that define that world. These figured worlds become cultural realms where "particular characters and actors are recognized, significance is assigned to certain acts, and particular outcomes are valued over others" (Holland et al., 2001, p. 53). Individuals' identities and the agency they carry both constitute and are constructed by and within those created cultural realms. In academic disciplines, figured worlds can comprise interwoven sets of disciplinary knowledge and values that are used to assess each other and novices in their acts to seek entrance to or ongoing acceptance within those worlds (Rahm, 2007). Those specialized worlds are also always multiple and alternate: intersecting to varying degrees but never completely isolated. In any everyday or specialized experiences our identities and their figured worlds collide. For example, figured worlds of engineering education intersect with figured worlds of science, technology, university life, youth culture, being and becoming gendered and more (Tonso, 2006). Those intersections and contradictions are also sites of meaning making and agency, changing the nature of disciplinary educational experiences (Francis, 2012). "The space of freedom that is the space of play between these vocations is the space of the author" (Holland et al., 2001, p. 238).

Figured worlds are both durable and changeable. They are socially reproduced through ongoing communities and yet they are made and re-made through boundary and identity work. Rahm and Moore (2016) for example spotlight the ways in which students' identities-in-practice in an informal STEM program create newly figured worlds of personal scientific participation, different from the formalized worlds of school and postsecondary science. That process of new figuration can be aided through the invocation or creation of counter-worlds: a world that defines only what this one is not (Holland et al., 2001). Political speeches are strong examples where a "world we don't want" is essential to the discursive creation of the world that the candidate proposes.

These conceptions of alternate, intersecting and counter-worlds along with the framing of identities as social works in progress bare a strong coherence with concepts of boundary work. Emerging from the sociology of science, Gieryn (1983, 1999) worked to reframe the problem of demarcation (i.e., distinguishing science from non-science) from a theoretical and sociological challenge to a practical and ongoing element of scientific work. Taking inspiration from the constantly remade social and political boundaries that shape cartographies, he labelled as "boundary work" the continuous acts of figured world creation that scientists engage in when they frame their work through what it is not (e.g., "not-religion" "not-mechanics" "not this kind of research but that kind"). What is striking is the way that boundary work becomes not the work of peripheral participants, of novices seeking entry into a social world, but an act that always exists at the core of a figured world. It is central to its meaning, and therefore an essential aspect of the creation of those worlds and all participants' experiences in them.

Rahm, Miller, Hartley and Moore (2003) examine a similar conceptualization in attempting to identify the meaning of scientific authenticity within a formal/informal partnership between STEM students, teachers and scientists. Their analysis breaks from the notion that disciplinary cultures must only be transferred from expert to novice, that authentic practice always begins as peripheral participation through gradual learning or simulation of disciplinary practices. They conceptualize scientific disciplinary authenticity as an emergent property, meaning-making that unfolds through negotiation between all of the participants.

Where multiple social worlds intersect, boundary objects can also become important features for negotiating and navigating between and across them. Boundary objects can be tangible (e.g., maps, Star, 2010), textual (e.g., science news stories, Polman & Hope, 2014) or conceptual (e.g., the ecological meaning of resilience, Brand & Jax, 2007). Their primary feature is that they are meaningful in multiple social worlds, even though those definitions may be different or even contradictory. Collaborations between experts embedded in disciplinary practices can proceed, even in the absence of shared understanding and language, when boundary objects can be acted upon in ways that are meaningful to each social world (Gorman, 2002; Star, 2010). Similarly, Holland et al. (2001), following Vygotsky (1967), highlight the ways that various objects can be pivots that allow individuals entry into new and emerging and figured worlds. While they focus on the value of objects for novices, we suggest that objects of various types can also become inter- and intra-subjective boundary objects, allowing individuals (novice and expert) to move back and forth, and through inter-spaces, between different figured worlds in a way that is analogous to expert communities tacking between various boundary object meanings. They may also allow designer and visitor to entrance into and participation in each other's figured worlds.

And in that sense, boundary work can also be boundary play. Holland et al. (2001) begin with Vygotsky's notions of play in children as an entry point in understanding their roles in the various games they play in their social lives. They extend it, however, to recognize that it is a key process in any encounter with new, emerging or alternate figured worlds and counter-worlds. They focus on how people, through playful even sometimes contradictory interactions, come to create, share and participate in figured worlds. But more than that, they argue, even short term playful excursions into new, emerging and previously counter-worlds can dishabituate us from the active identities and figurations of our usual worlds, leaving players and their home worlds both transformed.

Public STEM and computing spaces, because of their multi- inter- and transdisciplinary character, offer a unique play opportunity in the boundary spaces between and among disciplines. Informal STEM educators, such as Rahm and Moore (2016) have already recognized the multiplicity that even the various overlapping and intersecting worlds of scientific practice create. Others have argued that the very idea of an integrated STEM education is itself a boundary object allowing educators to re-imagine the potential of intersecting worlds through collaborative STEM practice (Shanahan, Burke & Francis, 2016). Arguments over the meaning of STEM and its potentially integrative character have invigorated boundary work over what even constitutes STEM education (e.g., Weinstein, Blades, & Gleason, 2016). In contemporary scientific practice, driven often by large data sets and modelling (e.g., Michener & Jones, 2012) computing can also be understood as a world that not only exists as its own disciplinary culture and practice but also one that intersects with and integrates individual STEM disciplinary worlds (McLeod & Nerssessian, 2013, 2014). Similarly, computational thinking and modeling can bring together multiple disciplines within STEM (e.g., mathematics, physics and biology) in K12 classrooms, by enabling students and teachers to model multiple phenomena across different domains using similar modeling approaches that emphasize the underlying mathematical commonalities (Sengupta et al., 2015).

Just because multiple disciplinary worlds are possible within STEM and computing, however, it is not a given that play will happen; that is, the environment must facilitate and encourage it. As Carlone, Haun-Frank and Webb (2011) observed in classroom-based ethnographic studies, it is difficult to avoid falling into established patterns of identity and boundary work that re-create the figured world of disciplinary science in ways that reify the status quo. Public spaces, removed from established curricular trajectories can make room for different interactions, where play as an expert in alternative figured worlds may be transformative. Here we explore one such space.

## 3. The DigiPlay Learning Environment: Glass Box, Open Source, and Public

DigiPlay is a learning environment located in an indoor, public walkway at the University of Calgary. It consists of three 80" touch screens, each powered by a desktop. The screens currently display open source simulations of complex systems. A key scientific characteristic of these systems is *emergence*- i.e., larger scale patterns, such as flocks of birds and schools of fish, *emerge* from rather simple and relatively unplanned interactions between many individual entities (Holland, 1999; Wilensky & Resnick, 1998). Visitors can use the touch-sensitive screens to interact with simulated visualizations of complex systems in which the larger scale patterns (flocks) *emerge* as each virtual bird performs simple interactions with neighbouring birds. The simulations are programmed using the *Processing* programming language and visualization platform (Reas & Fry, 2007). *Processing* is open source, used by computer scientists and digital artists alike, and there is a strong online user community of experts and learners, making sample code and simulations created by them accessible to the public.

The *Processing* simulations we designed for DigiPlay are both open source and "glass box" (Wilensky & Reisman, 2006). The *open source* nature of the code makes it possible for visitors to interact with and modify the code that may have been originally created by an expert, and it also allows us as developers of DigiPlay to extend and modify functionalities of the Processing programming language itself, as needed. The *glass box* (du Boulay, O'Shea, & Monk, 1981) nature of Processing enables that visitors to access the underlying code, while the simulations are running in the form of dynamic visualizations in fullscreen mode, by simply hitting the "Escape" button once on the on-screen keyboard. DigiPlay visitors can directly interact with and modify the emergent patterns in the visualizations by adding new boids to the flock by touching the screens, and at the same time they can also make deeper changes to the way the individual boids interact by accessing the underlying code.

The algorithms we used are adapted from Reas & Fry's (2007) implementation of Craig Reynolds's classic algorithm for simulating flocking of birds (Reynolds, 1987). Each Boid (Reynolds termed each virtual bird a "Boid") in the simulation acts as a computational "agent", and the DigiPlay simulations can therefore can be understood as multi-agent simulations. The term "agent" here indicates individual computational objects or actors. It is the behaviors and interactions between these agents as well as elements of the environments in which they are situated, that give rise to emergent, system-level behavior (e.g., the formation and movement of a traffic jam or the spread of disease). Each agent in a multi-agent simulation makes its own decision. In Reynolds' algorithm, each Boid follows three simple rules: alignment, coherence and separation. "Alignment" means that a Boid tends to turn in the same direction that nearby birds are moving. "Separation" means that a Boid will turn to avoid another bird which gets too close. "Cohesion" means that a Boid will move towards other nearby birds. Therefore, the emergent patterns represented in the simulations, i.e., flocks, do not result from averaging over the entire population, but from the aggregation of the outcomes of individual-level decisions of many agents. Multi-agent simulations can therefore enable the users to understand and explain complex, aggregate-level, emergent patterns in terms of the local, individual-level rules of interaction. Pedagogically, this is an important affordance of multi-agent systems, as we explain next.

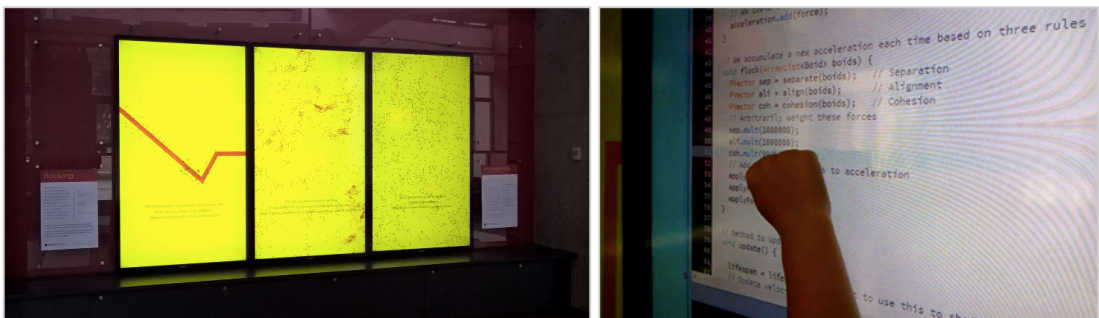

*Figure 1*. The Digiplay environment (left) and a young visitor hacking the DigiPlay code (right)

Our choice of multi-agent simulations as the means to simulate complexity is based on prior research illustrating that while learners at all levels find understanding emergent processes challenging (Chi, 2005), multi-agent-based computational models and simulations can be successful in helping them overcome the difficulties (Dickes, Sengupta, Farris & Basu, 2016; Danish, 2014; Wilensky & Resnick, 1999). These studies show that curricula that utilize agent-based models can help students understand complex systems and emergence by grounding emergent phenomena in terms of their embodied, agent-level intuitions. The ability to fluidly and meaningfully move, to "dive in" and "step out" of emergent phenomena (Ackerman, 1996) as afforded by the touch screens and glass box features, enables learners to connect their agent-level, embodied and intuitive understanding to the emergent-level outcomes (Wilensky & Reisman, 2006; Danish, 2014; Dickes et al., 2016). But additionally, it allows them access and play in multiple figured worlds, both the imagined world of the boids and the social worlds they inhabit physically and, through the code, virtually.

Finally, the *public* nature of the space ensures that anyone can walk in and interact with DigiPlay. The users can access just-in-time information as they are interacting with the touch screens, which provide them instructions for modifying the live simulations. The "rules" obeyed by the Boids are also explained in the form of posters on glass walls surrounding the monitors. In addition, there is often an on-site facilitator present to provide the public direct and live access to expertise. The on-site facilitator is one of the members of the team that developed the exhibit (including the first author), and they have a deep understanding of the underlying code. The facilitator's primary role is encouraging the visitors to "hack" the simulations, by showing them how to access the underlying code, and pointing them to relevant areas in the code that can be easily altered to potentially powerful effect.

Our work bears similarities to Seymour Papert's vision of Brazilian samba schools as a desirable image of learning with computational media. He envisioned "computational samba schools" where people would come to learn through and about technology in a self-motivated, community-supported fashion (Papert, 1980). Zagal and Bruckman argued that "computer clubhouses", after-school drop-in computer centers of the Intel Computer Clubhouse Network typically housed in inner city neighborhoods in the US (and later, worldwide), should be understood as examples of such computational samba schools (Zagal & Bruckman, 2005). In computer clubhouses, the intention is to situate learners in *communities of practice* (Lave and Wenger, 1996), in the process transforming technological participation from the act of learning coding in isolation to one of creating a public artifact that is valued by a broader community. Similarly, while it is a short term engagement, when visitors interact with code in DigiPlay they are interacting with code that is valued by a broader community of scientists and digital artists. The contrast, however, is that when visitors in DigiPlay directly interact with and modify artifacts created by experts they do so in a truly open and public space rather than in a learner's or apprentice's space. Thus, not only do visitors create public artifacts, but coding in DigiPlay is also a truly public experience.

## 4. Methods

The data are in the form of field notes, video recordings and photographs of visitor interactions with DigiPlay. The observed conversations reported here took place among the visitors, and are typically their interpretations of the code and the patterns of flocking behavior displayed in the simulations as they are modified.

We report our analysis in the form of a retrospective observational case study of a single visit by a group of three visitors. We chose this case from a corpus of over 20 cases, where each "case" corresponds to a "visit" either by an individual or a group of visitors. These visits lasted on an average 10-15 minutes and were recorded through short audio and/or video clips filmed with a mobile phone. Because DigiPlay is housed within a university, many of these cases involve visitors who were likely affiliated with the university, such as graduate and undergraduate students. However, this case was chosen because the visitors were two young school-aged boys (Sam and Rex, pseudonyms), who visited DigiPlay along with an adult (Mary, pseudonym). Their age suggested that their figured worlds of science, programming and technology might all be emergent and that the phenomenon of play may be central to their everyday interactions with the world.

We adopted a phenomenographic approach for our analysis. Our choice of phenomenography was based on Marton's (1981) argument that phenomenography deals with the forms of immediate experience as well as conceptual thought and physical behavior (Marton, 1981, pp 41 - 42). This is particularly important for our theoretical focus on figured worlds and boundary work, which involves not only how we act in the world, but also how we conceptualize and interpret our actions and the environment where we *are* and *might be* situated. Our analysis is grounded in a "nondualist ontology" (Marton & Booth, 1996): "there is only one world, a real existing world that is experienced and understood in different ways by human beings; it is both objective and subjective at the same time. An experience is now a relationship between object and subject that encompasses them both" (p, 537). An object or an event, in this view, is also "seen" as the *emergent* phenomenon - i.e., "the complex of all different ways it *might be* experienced" (Marton & Booth, 1997, p 113). The figured worlds of the visitors are also emergent and dynamically constructed. Visitors' figured worlds are evident to us, albeit interpretively, through visitors' actions on the computational elements in DigiPlay, as well as their interactions with others in the space.

Our analysis is thematic in nature. We rely on observations of the visitors' actions and conversations, facilitated by the recordings and field notes, as our primary source of data. To look for instances of boundary work and boundary play, we focused on the actions the visitors undertake in DigiPlay as they alter the visualizations and the code.

# 5. Findings

A common observation across most visitors in DigiPlay is the contrast between the original intent of particular elements of the code from the perspective of the exhibit designers on one hand, and the modifications carried out by the visitors on the other. In our analysis, fragments of the underlying code therefore serve as *boundary objects* between the *figured worlds* of the exhibit designers and the visitors. The figured worlds of the visitors, however, are malleable, and so are the boundary objects. With the passage of time, visitors' actions on the code are further shaped by the conversations that unfold in the space among the visitors, and vice versa. We chose this case because it illustrates vividly the complex nature of the visitor experience.

*5.1. Code and Flocks: The Designer's Boundary Objects and Figured Worlds*

Sam and Rex began their interactions with the simulations by touching the screens to add new Boids to the simulations, and Mary began reading aloud the "rules" obeyed by the Boids from one of the posters on the DigiPlay glass panels. After a couple of minutes, the facilitator pointed to a Boid on screen and explained to Sam and Rex that the Boids follow three concurrent rules with their nearest neighbors: alignment (turn toward the nearest Boid), coherence (move closer to the nearest Boid), and separation (maintain a minimum separation with the nearest Boid). His explanations were both verbal and gestural: he explained out loud and demonstrated through embodied movements how his body would turn toward (align), move toward (cohere) and separate from the nearest Boid. His intention, in doing so, was to explain how larger scale patterns, such as a flock of several Boids, can emerge from local, individual-level rules obeyed by each Boid, showing a strong understanding of the scientific purpose of the simulations.

The facilitator then asked Sam and Rex if they would like to hack the simulations. Sam expressed disbelief, and ran away to a chair and feigned to pass out, lying down in the chair. Rex was also shocked, but both of them were excited, as evident in their jumping up and down. At this point, the facilitator pointed them to the portion of the code where they could control the relative weights of the three forces: alignment, coherence and separation (Figure 2, left). The facilitator, who also designed the code for the simulations, pointed out that the three forces were weighted as follows: separation (1.6), alignment (1.5), and coherence (1.0).

It had taken the first author several weeks to optimize these parameters at these values, through repeated tests, given his objective of simulating flocks of birds. So, this was the exhibit designer's figured world of this exhibit: the code, and the resultant simulation, should realistically depict how flocks of birds form. Figured worlds rely on cultural artifacts that serve as boundary objects (Holland et al., 2001; Cole 1996), and the cultural artifact in this case was the open source code on which this simulation was based. As Cole (1996) points out, cultural artifacts have developmental histories, which assume both an obvious and necessary material object as well as an ideal or a conceptual aspect (intentionality). In this case, the developmental history corresponds to the multiple versions of the algorithm and the code--authored by other experts and available in open source format--that the designer compared with and adapted from. Adaptations from the original and related versions were necessary for several reasons, the primary being the geometry of the screen and the size of the Boids. As in any multi-agent simulation, these factors greatly altered the movement and density of the Boids, and hence the emergent patterns or flocks. The adaptations were reflected in the choice of the numbers that act as relative weights of the three forces, and are an example of the "substance" of intentionality (Holland et al., 2001; pp 61) that was embedded in the figured world of intended use (as a realistic simulation) from the perspective of the designer.

*5.2. Boid Prisons and Code Use: Visitors' Figured Worlds & Boundary Work*

For Rex and Sam, it was a lot less important to simulate flocking birds. Once they realized that they could easily make substantial changes to the emergent behaviors by altering the relative weights of the three forces, they began changing them to arbitrarily high values: separation (2300), alignment (1620), and coherence (2073). Once they made the changes and ran the simulation, Rex explains to Sam: *"They [Boids] have to come together but they also have to move far away."* Standing behind them, Mary points to the Boids getting close to each other and then turning away, and exclaims, *"Oh look they are bonking."* At this point, Rex starts tapping the screen toward the bottom left corner, and adds a lot of Boids (~one hundred), and it results in a "blob" of Boids. Sam exclaims and laughs, so does Mary; Sam says, *"I love this"*. Sam then says that it looks a "glitch" to him - meaning that the Boids crowding together (instead of flocking) looked as if he made a mistake in the code. This contrast is shown in Figure 3. At this point, Mary interjects: *"Look what you have done to them - they cannot escape… I am gonna call them the "swarm""*. Sam then interjects: *"I am gonna call them the 'prison'"*. Rex then says, *"Anyone who gets close to the prison, gets sucked in it"*. Mary says reflectively: *"Hmmmm"*. The visitors are all now looking intently at what happens when Rex keeps adding new Boids very close to the blob. The new Boids do indeed appear to stick to the blob, and the blob keeps growing in size, and almost no Boid escapes the blob. After several seconds of intent observation and adding new Boids, Rex explains: *"I guess they have to stay as close as possible and they have to stay away as far as possible… the forces of being closer are stronger"*. Mary says: *"I guess they are cohering, right? Is that what is happening?"* Sam interjects: *"The jail cell"*. Mary now responds: *"Huh.. jail cells.. So do you think that this model could be used to explain what would happen..umm.. in .. overcrowding in prisons?"* Sam responds: *"yeah!"*. Mary then smilingly explains to the facilitator that she always tries to "sneak in" lessons on social justice.

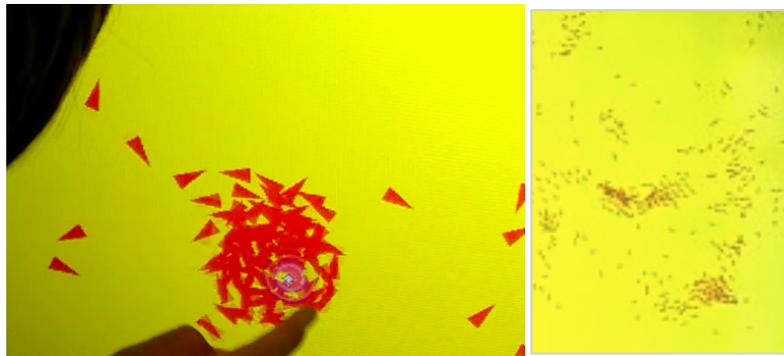

*Figure 3*. Rex and Sam's Boid Prison (left) and an expert's implementation of "Flocking" (right)

A key characteristic of boundary objects is their ill-structured nature, allowing them to have different meanings in the different social worlds that they cross and therefore to be acted on in completely different ways (Star, 2010). This is particularly relevant in situations of cooperation without consensus, or in asynchronous engagements by individuals inhabiting different figured worlds because engagement with the objects becomes dominated by localized meanings. The visitors did indeed localize (i.e., appropriate) the code and the simulation, both by making simple but powerful changes to it and through their interpretations of the emergent patterns in the simulation. Their figured world of the meaning and purpose of the simulation (and perhaps of simulations more broadly) was distinct from that of the designers. However, the code here also serves as legitimate boundary object from the designers' perspective in terms of supporting the intended purpose of helping the visitors learn about emergence. This becomes evident in the form of visitor's explanations of the emergent patterns in terms of agent-level rules. See for example, Rex's explanations of why the Boids are "bonking" and Mary's explanation of why prisons are forming, in the earlier paragraph, that exemplify this claim. Developing such multi-level explanations is the central goal of using multi-agent simulations to model complex systems (Wilensky & Reisman, 2006; Dickes et al., 2016). At the same time, the visitors' figured worlds of Boid Prisons were co-constructed through joint action on the code and interpretations of how the code alterations were affecting the simulated visualization. In contrast, these figured worlds - Boid prisons, jail cells, and a cluster of Boids as a model of overcrowding - were dissonant from the figured world (bird flocking) of the exhibit designer. Stepping back, the code may have served as a boundary object allowing play between the larger disciplinary cultures of natural and social sciences in which the designer's and visitors' (respectively) figured worlds of the simulation were embedded.

## 6. Discussion

So how can public computation alter the status quo of STEM experiences and education? We argue that coding as public experience can be understood as the coming together of canonical disciplinary practices and private interpretations as the public engages in the playful boundary work of generating and configuring multiple figured worlds. In the world of science, theories can be understood as figured worlds; and as Pickering (1996) argued, conceptual development in science is deeply intertwined with the development of representational practices. We see a striking parallel in the experience of visitors: the configuration and reconfiguration of their figured worlds is shaped by a deeply intertwined relationship between their coding and their interpretations of emergent simulations. The open source code enables public access to archived expertise, and the public nature of the environment invites and validates multiple figured worlds that would not otherwise coexist in a traditional classroom or learning environment.

## 10. Acknowledgements

We thank Sharon Friesen and Dennis Sumara for their vision and support for this project, and Imperial Oil Foundation for funding support.